\documentclass[aps,pra,twocolumn,showpacs]{revtex4}  

\usepackage{epsfig}
\usepackage{graphicx}  
\usepackage{bm}        
\usepackage{amssymb}   
\usepackage{amsmath}

\begin{document}


\title[Resonance in a driven two-level system]{Resonance in a driven two-level system: analytical results without the rotating wave approximation}
\author{Yu.~V. Bezvershenko and P.~I. Holod}
\affiliation{National University of Kyiv-Mohyla Academy, 04070 Kyiv, Ukraine}
\affiliation{Bogolyubov Institute for Theoretical Physics, 03680 Kyiv, Ukraine}
\email{yulia.bezvershenko@gmail.com,holod@ukma.kiev.ua}

\begin{abstract}
We consider the problem of two-level system dynamics induced by the time-dependent field ${\bf B}=\{a(t)\cos\omega t,a(t)\sin\omega t,\omega_0\},$ with $a(t)\propto\textmd{cn}(\nu t,k).$ The problem is exactly analytically solvable and we propose the scheme for constructing the solutions. For all field configurations the resonance conditions are discussed. The explicit solutions for $N=1,~2$ we obtained coincide at $\omega=0$ in the proper parameter domain with predictions of the rotating wave approximation and agree nicely with numerical calculations beyond it.
\end{abstract}


\pacs{02.30.Gp, 03.65.-w, 76.20.+q}

\maketitle

\section{Introduction}

Dynamics of model two-level systems in the external fields is of fundamental interest, as this problem appears in various physical systems. They are, for instance, spin 1/2 in the magnetic field \cite{Rabi}, two-level atom interacting with classical field \cite{quantopt}, driven artificial two-level systems in superconducting Josephson devices \cite{JosC}. As well this problem is significant for physical realization of quantum gates \cite{qc}. But still there are few time-dependent field configurations for which the problem is exactly analytically solved \cite{Bambini,Jha,Xie}, so in many cases the solutions are given using perturbation theory or computed numerically \cite{Shirley,Barata,Frasca}.

One of the physically significant fields is linearly polarized (LP) harmonic wave. In particular, it is used in magnetic and optical resonance experiments \cite{Rabi,quantopt}. There is a number of approximate approaches to this problem. In particular, when one assumes the amplitude of the oscillating field to be small in comparison with the static field along the third axis and the frequencies of these fields are near-resonant. In this case the rotating-wave approximation (RWA) is valid and one can easily get the solutions known as Rabi oscillations \cite{Rabi}. The understanding of magnetic resonance, meaning the resonance conditions, is based on these approximative treatment, and, for example, the resonance frequency is given by series in the amplitude (the first correction is called Bloch-Siegert shift) \cite{Bloch_Ziegert,BlSie3}. On the other hand, not so long ago this problem was shown to be exactly analytically solvable in terms of Heun functions \cite{Xie}. But as the apparatus of the latter is not developed enough, it is quite difficult to extract physical information from the solutions.

In this Letter we consider the generalization of the Rabi field which at different values of parameters corresponds to several physically significant situations: nonlinearly modulated Rabi field, N-soliton pulse and the linearly polarized field, which approximates harmonic one. We succeed to obtain explicit analytical expressions describing dynamics of a two-level system in the proposed time-dependent field and, consequently, to get conditions of resonance in it. The emphasis should be laid on the third field configuration, as it reflects the idea to replace the cosine potential with another one resembling it with arbitrary accuracy but corresponding to the exactly solvable problem in terms of well-developed functions. The exact results we obtained coincide at small amplitude of the field and near-resonant frequencies with predictions under the RWA, while beyond it they are in agreement with numerical calculations of the problem with the linearly polarized harmonic field. The resonance conditions in the analytic form for this field, in particular, reproduce at small field amplitude the Bloch-Siegert shift.

\section{Formulation of the problem}
We consider the problem of a two-level system dynamics in the external field
\begin{equation}\label{1}
    {\bf B}(t)=\{2 a(t)\cos\omega t,2 a(t)\sin\omega t,\omega_0\},
\end{equation}
where $a(t)=Nk\nu \textmd{cn}(\nu t,k),$ $N\in\mathbb{Z}.$ Here $\textmd{cn}(\nu t,k)$ denotes the Jacobi (cnoidal) elliptic function of modulus $k,~0< k< 1$ \cite{WW}.
Let ut analyze briefly the range of physical situations given by the field ${\bf B}(t)$  \cite{RemarkEberly}. For the arbitrary parameters it can be understood as nonlinearly modulated Rabi field, meaning the nonlinear function $a(t)$ rather than constant amplitude $a$ in ${\bf B}_R(t)=\{2a\cos \omega t,2a\sin\omega t,\omega_0\}$ \cite{Rabi}. In the case $k=1$ the amplitude is $a(t,k=1)=N\nu\textmd{sech} \nu t,$ so the field is rotating with the $N$-soliton envelope \cite{Bambini}. The third situation occurs under $\omega=0:$
${\bf B}_{LP}(t)=\{2 a(t),0,\omega_0\}.$ If the limit $N\rightarrow\infty,~k\rightarrow0$ ($Nk=const$) is taken, this field is exactly the linearly polarized harmonic wave. There is also an approximate limit, valid for appropriately small $k,$ when $\textmd{cn}(\nu t,k)\approx\cos(\frac{\pi}{2K(k)}\nu t),$ here $K(k)$ is the complete elliptic integral of the first kind. As the period of the cnoid is $4 K(k),$ due to definition of the frequency we obtain $\Omega=\nu\pi/[2K(k)]$ as an "effective frequency" of the cnoidal field. Therefore, finding the dynamics in the field (\ref{1}), setting $\omega=0$ and employing the range of small enough $k$ gives solution to the problem with the field $\{\cos\Omega t,0,\omega_0\}.$

The Hamiltonian of our problem has the form:
\begin{equation}\label{2}
    \hat{\mathcal{H}}={\bf B}(t)\cdot\hat{{\bf \sigma}},
\end{equation}
with $\hat{{\bf \sigma}}=\{\hat{\mathcal{\sigma}}_i\}$ - Pauli matrices. Let $|\pm\rangle$ be the eigenstates of operator $\hat{\mathcal{\sigma}}_3$ with
eigenvalues $\pm 1$, respectively. The evolution of a generic state of a two-level system $|\Psi(t)\rangle=\widetilde{C}_+(t)|+\rangle+\widetilde{C}_-(t)|-\rangle,$
where $|\widetilde{C}_+(t)|^2+|\widetilde{C}_-(t)|^2=1,$ is determined by the
Schr\"{o}dinger equation
\begin{equation}\label{5}
    \imath \hbar \frac{\partial}{\partial t}|\Psi(t)\rangle=\hat{\mathcal{H}}(t)|\Psi(t)\rangle.
\end{equation}
Considering the form of Hamiltonian (\ref{2}) for the given
choice of fields, we have the system of equations for amplitudes
$\widetilde{C}_{\pm}(t)$:
\begin{equation}\label{6}
\frac{\partial}{\partial t}\left[
                                            \begin{array}{c}
                                              \widetilde{C}_+(t) \\
                                              \widetilde{C}_-(t) \\
                                            \end{array}
                                          \right]
=\frac{1}{2\imath}
                                          \left[\begin{array}{cc}
                   \omega_0 & a(t)e^{-\imath \omega t} \\
                   a(t)e^{\imath \omega t} & -\omega_0 \\
                 \end{array}\right]
                 \left[
                                            \begin{array}{c}
                                              \widetilde{C}_+(t) \\
                                              \widetilde{C}_-(t) \\
                                            \end{array}
                                          \right].
\end{equation}
We will choose the initial conditions as \begin{equation}\label{incond}\widetilde{C}_+(0)=1,\quad\widetilde{C}_-(0)=0,\end{equation} meaning that the dynamics starts from the $|+\rangle$-eigen\-state.

\section{Scheme for constructing the solutions}
Introducing new variables $C_\pm(t)=\textmd{e}^{\pm\imath\omega_0 t/2}\widetilde{C}_\pm(t)$ the system of Eqs. (\ref{6}) can be reduced to independent equations of the second order. In terms of the dimensionless variables they take the following form
\begin{equation}\label{centreq}
\frac{d^2 C_{\pm}}{d\tau^2}+\left[\frac{\textmd{sn}\tau~\textmd{dn}\tau}{\textmd{cn}\tau}\pm\imath\Delta\right]\frac{d C_{\pm}}{d\tau}+N^2k^2\textmd{cn}^2\tau~C_{\pm}=0,
\end{equation}
where $\textmd{sn}\tau,~\textmd{cn}\tau,~\textmd{dn}\tau$ are the Jacobi elliptic functions of mo\-du\-lus $k$, $\tau=\nu t$ and $\Delta=(\omega_0-\omega)/\nu$ is dimensionless detuning.
After the change of variable $s=\textmd{sn}(\tau,k)$ the equations (\ref{centreq}) are
\begin{equation}\label{centreqs}\frac{d^2 C_{\pm}}{ds^2}-\left[\frac{k^2s}{1-k^2s^2}\pm\frac{\imath\Delta}{\sqrt{w^2}}
\right]\frac{d C_{\pm}}{ds}+\frac{N^2k^2}{1-k^2s^2}C_{\pm}=0,
\end{equation}
thus they can be classified as the equations of the Fuchs type \cite{Ince} on the Riemann surface corresponding to the algebraic curve $w^2=(1-s^2)(1-k^2s^2)$ \cite{WW}.

Two obvious limiting cases ($\Delta=0$ and $k=1$) lead to simplification of the Eqs. (\ref{centreqs}) to the hypergeometric type which could be treated easily and are physically meaningful.
In the case $\Delta=0,$ which means the resonance between frequencies of constant and oscillating fields, the Eq. (\ref{centreqs}) is simplified to equation of hypergeometric type, with solutions:
\begin{equation}\label{tch}
C_{+}=\textmd{T}_N(-ks),~C_{-}=-\imath \sqrt{1-k^2s^2}\textmd{U}_{N-1}(-ks)\end{equation}
where $\textmd{T}_N(x)$ and $\textmd{U}_N(x)$ are Chebyshev polynomials of the first and second kinds, respectively. The amplitudes satisfy proper initial condition automatically.
Other situation can be obtained setting $k=1$, so that initial field is reduced to the rotating field with N-soliton envelope $a(t,k=1).$ The equations are also hypergeometric and the solution is
\begin{equation}\label{Jacobi}C_{+}=P_N^{(a,b)}(s),\end{equation}
where $P_N^{(a,b)}$ is the Jacobi polynomial with complex parameters $a=-(1-\imath\Delta)/2,$ $b=-(1+\imath\Delta)/2$. Here it is important to mention that the initial condition should be substituted by $C_+(-\infty)=1,~C_-(-\infty)=0,$ as the field is non-periodic and its action starts at $t=-\infty.$

Let us seek for solution of the equation (\ref{centreqs}) when "plus" is chosen in the form
\begin{equation}\label{CAnsatz}
    f_{+}=\exp\left[\int_0^s\left(\mathcal{R}_0(s')+\imath\frac{\mathcal{R}_1(s')}{\sqrt{w^2}}\right)ds'\right],
\end{equation}
where $\mathcal{R}_0(s)$ and $\mathcal{R}_1(s)$ are rational functions in $s.$ As a result, the initial equation (\ref{centreqs}) comes up to the system of Riccati-type equations
\begin{subequations}
\begin{eqnarray}
\label{R0ei}
&\mathcal{R}_0'+\mathcal{R}_0^2-\frac{k^2s}{1-k^2s^2}\mathcal{R}_0+\frac{(\Delta \mathcal{R}_1-\mathcal{R}_1^2)}{w^2}+\frac{N^2k^2}{1-k^2s^2}=0,\\ \label{R1}
&\mathcal{R}_1'+\mathcal{R}_1(2\mathcal{R}_0-\frac{s}{s^2-1})-\Delta \mathcal{R}_0=0.
\end{eqnarray}
\end{subequations}

Obviously, $\mathcal{R}_0(s)$ in terms of the amplitude modulus has the form
\begin{equation}\label{R0}
\mathcal{R}_0(s)=\frac{1}{2}\frac{d}{ds}\ln|f_+|^{2}.
\end{equation}
Since solutions in both limiting cases imply that $|C_+|$ is a polynomial function, we will try $f_+$ in the form $$|f_{+}|^2=\mathcal{P}_{2N}(s,e_i)=\prod_{i=1}^N(s^2-e_i^2),$$ where $\mathcal{P}_{2N}$ is a polynomial of the 2$N$-th order with roots $e_i.$ Then due to (\ref{R0}) $\mathcal{R}_0(s)$ can be easily calculated.
The solution of (\ref{R1}) with known $\mathcal{R}_0(s)$ is given by
$$\mathcal{R}_1(s)=\Delta\int_0^s e^{-\int_y^s\left[2\mathcal{R}_0(x)-\frac{x}{x^{2}-1}\right]dx}\mathcal{R}_0(y)\textmd{d}y.$$
The equation (\ref{R0ei}) serves as the consistency condition and defines the roots $e_i.$ With known $\mathcal{R}_0(s)$ and $\mathcal{R}_1(s)$ one can obtain the expression for the $f_+$ due to (\ref{CAnsatz}). Analogically the solution $f_-$ of (\ref{centreqs}) standing for the "minus" sign can be found. Then one needs to chose two linearly independent solutions of (\ref{centreqs}) and create linear combinations satisfying initial conditions (\ref{incond}).

Using this scheme we can find the solutions for $N=1$
\begin{subequations}\label{N1}
\begin{eqnarray}
\textmd{C}_+=
\varepsilon_1\sqrt{s^2+e_1^2}~\textmd{e}^{-\imath \varphi_1}+\varepsilon_2\sqrt{-s^2-e_2^2}~\textmd{e}^{-\imath \varphi_2},\\
\textmd{C}_-=-\imath\varepsilon_2\sqrt{s^2+e_1^2}~\textmd{e}^{\imath \varphi_1}+\imath
\varepsilon_1\sqrt{-s^2-e_2^2}~\textmd{e}^{\imath \varphi_2},
\end{eqnarray}
\end{subequations}
where
$e_{1,2}^2=\left[\left(\Delta^2-1\right)\pm\sqrt{\left(\Delta ^2-1\right)^2+4 k^2 \Delta ^2}\right]/(2 k^2),$\\
$\varepsilon_i=\sqrt{(-1)^{i+1}e_i^2}/(e_1^2-e_2^2),~i=1,2.$ Here the time-dependent phase
$$\varphi_i=\Delta\left[\tau -(1+e_i^2)(e_i^2)^{-1}~
\Pi\left(-(e_i^2)^{-1};\textmd{am}(\tau,k),k\right)\right]$$ is defined in terms of the incomplete elliptic integral of the third kind \\ $\Pi\left(n;\textmd{am}(\tau,k),k\right)=\int_0^s[(1-nx^2)\sqrt{(1-x^2)(1-k^2x^2)}]^{-1}dx$ and $\textmd{am}(\tau,k)$ denotes the Jacobi amplitude function \cite{WW}.

Increase of $N$ (physically it corresponds to the increase of the field amplitude) results in the complication of expressions for the amplitudes but doesn't change its structure.
Explicit solutions for $N=2$ are the following
\begin{subequations}\label{N2}
\begin{multline}
\textmd{C}_+(\tau)=
\varepsilon_a\sqrt{(-s^2+e_1)(s^2-e_2)}\textmd{e}^{\imath \varphi(e_1,e_2)}\\+\varepsilon_b\sqrt{(-s^2+e_3)(-s^2+e_4)}\textmd{e}^{\imath \varphi(e_3,e_4)},\end{multline}
\begin{multline}\textmd{C}_-(\tau)=
\imath\varepsilon_b\sqrt{(-s^2+e_1)(s^2-e_2)}\textmd{e}^{-\imath \varphi(e_1,e_2)}\\-\imath\varepsilon_a\sqrt{(-s^2+e_3)(-s^2+e_4)}\textmd{e}^{-\imath \varphi(e_3,e_4)}.\end{multline}
\end{subequations}
Here $\varepsilon_a=\sqrt{-e_1e_2}/(e_3e_4-e_1e_2),~\varepsilon_b=\sqrt{e_3e_4}/(e_3e_4-e_1e_2),$\\ $e_{i}=\left[-4(\Delta ^2-9)+(-1)^{i+1}\sqrt{2}A_\pm\right]/(36 k^2),~i=\overline{1,4},$\\
$A_\pm=\left[9 \left(9\pm\sqrt{d}\right)+54 \left(1-2 k^2\right) \Delta ^2-7 \Delta ^4\right]^{1/2},$
where "plus" stands for $e_{1,2}$
and "minus" for $e_{3,4},$ $d=\left[\left(\Delta ^2-9\right)\left(\Delta ^2-1\right)+12(\Delta k)^2\right]^2+\left(\Delta ^2-9+48 k^2\right)^2$ $\times(4\Delta k)^2$
and the phase \[\varphi(e_l,e_n)=\Delta \left[2 \tau/3 -\gamma_l\Pi_l+\gamma_n \Pi_n\right],\]
$\gamma_{l(n)}=(e_{l(n)}-1)(-4+e_{l(n)}+3e_{n(l)})/\left[3e_{l(n)}(e_1-e_2)\right],$ where $(l,n)$ stands for the pairs $(1,2)$ and $(3,4),$ and
$\Pi_l=\Pi\left(\textmd{e}_l^{-1};\textmd{am}(\tau),k\right).$

\section{Resonance conditions}
If the amplitudes are known, one can easily calculate the expectation values of spin momentum projections which are defined as
$\textmd{S}_i(t)=\langle\Psi(t)|\hat{\sigma}_i|\Psi(t)\rangle.$ They stand for the Bloch vector vector ${\bf S}(t)$,
whose evolution takes place on sphere. Its third component $S_3(t),$ which is useful for physical interpretations, can be found in terms of the amplitudes by the expression
\begin{equation}\label{10}
\textmd{S}_3(t)=|\widetilde{C}_+(t)|^2-|\widetilde{C}_-(t)|^2,
\end{equation}
where the initial conditions (\ref{incond}) are taken into account.

The problem of two-level system dynamics in the Rabi field ${\bf B}_R,$ which approximates the linearly polarized harmonic field when the amplitude of latter is small
and the frequencies are near-resonant (RWA), can be easily solved \cite{Rabi}. The third component of the Bloch vector
oscillates (Rabi oscillations) due to
\begin{equation}\label{12}
S_3^{(R)}(t)=[\widetilde{\Delta}^2+4a^2\cos\Omega_R t]/\Omega_R^2,
\end{equation}
where $\Omega_R=\sqrt{\widetilde{\Delta}^2+4a^2}$ is the Rabi frequency and
$\widetilde{\Delta}=\omega-\omega_0$ is the detuning between the rotating and the
constant field frequencies.

The term \emph{resonance} we will refer to the situation when under some special relations involving characteristic frequencies, system evolves from $|+\rangle$ to $|-\rangle$ state, thus minimum of $S_3(\tau=\tau_0)$ is $-1.$ In the Rabi field ${\bf B}_R$ the only resonance condition is $\omega=\omega_0,$ then the system performs oscillations with the period $T_R=\pi/a$ between states. The situation when the characteristic period of the dynamics is $T_R$ we will designate as the \emph{Rabi resonance}.

\begin{figure}[ht]
\includegraphics[width=\linewidth]{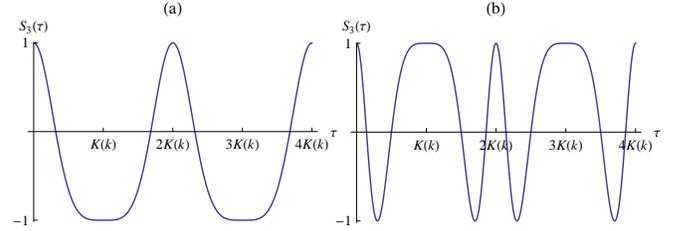}
\caption{\label{pict1}Resonance in the nonlinearly modulated Rabi field.
(a) $N=2,~k_{res}=0.7,$ (b) $N=3,~k_{res}=0.89.$}
\end{figure}

In the \emph{nonlinearly modulated Rabi field} $\omega=\omega_0$ is not enough to satisfy $S_3(\tau_0)=-1.$ Using the explicit expressions for the amplitudes given by (\ref{tch}) we obtain
$$S_3(\tau)=2\textmd{T}_N(-k\textmd{sn}\tau)^2-1.$$ Thus $\textmd{T}_N(-k_{res})^2=0$ is the additional resonance condition (Fig.~\ref{pict1}). For the situation of the \emph{N-soliton pulse} ($k=1$) the only resonance condition is $\nu=\omega_0.$

\begin{figure}[ht]
\includegraphics[width=\linewidth]{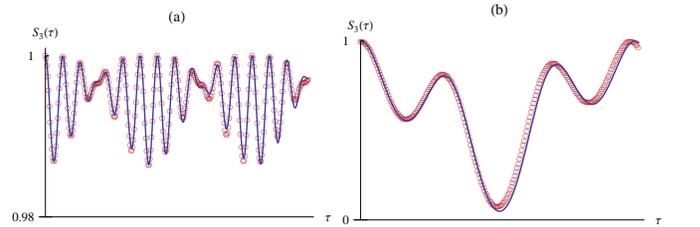}
\caption{\label{pict01}Comparison of the exact result for $N=2$ (solid line) in the field ${\bf B}_{LP}(\tau)$ with the numerical integration (circles) of the problem with the linearly polarized harmonic field $\{2a \cos{\tau},0,\omega_0\},$ $a=2k.$ Here $k=0.25,~a=0.5,$ (a) $\Delta=0.4,$ (b) $\Delta=12.$}
\end{figure}

In the case of two-level system driven by ${\bf B}_{LP}(\tau),$ $S_3(\tau)$ can be calculated for $N=1,2$ using explicit expressions (\ref{N1})-(\ref{N2}). Our analytic results are in good agreement with numerical calculations for the linearly polarized harmonic wave in wide range of parameters (see Fig.~\ref{pict01}).
As the field ${\bf B}_{LP}(\tau)$ has two characteristic frequencies: $\omega_0$ and $\Omega,$
the expected resonance condition is
\begin{equation}\label{BlSie_gen}\omega_0=\frac{\pi}{2K(k)}\nu.\end{equation}
For small $k$ the last expression can be expanded as series
\begin{equation}\label{BlSie}\omega_0=\nu(1-\frac{k^2}{4}-\frac{5k^4}{64}-...).\end{equation}
For $N=1,$ $k$ plays the role of the constant amplitude $a$ in the corresponding ${\bf B}_R,$ thus (\ref{BlSie}) coincides with expression for the Bloch-Siegert shift \cite{Bloch_Ziegert,BlSie3}.
\begin{figure}[ht]
\includegraphics[width=\linewidth]{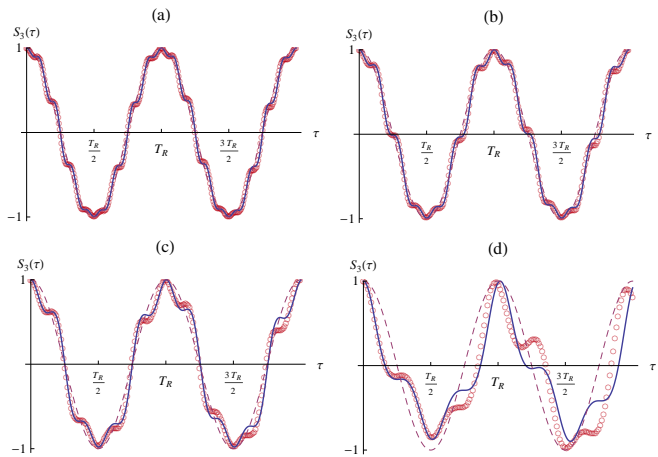}
\caption{\label{pict2}Comparison of the exact result for $N=2$ (solid line) in the field ${\bf B}_{LP}(\tau)$ with the numerical integration (circles) of the problem with the linearly polarized field $\{2a \cos{\tau},0,\omega_0\},$ $a=2k.$ Here $\nu=\omega_0$ and $k$ are chosen to be solutions of the $k~_{2}F_1(1/2,1/2;1;k^2)=(2m)^{-1},$ therefore, the Rabi oscillations (dashed line) with period $T_R=2\pi/a$ under small $k$ are expected to occur. (a) $k=0.1245,$ (b) $k=0.1655,$ (c) $k=0.2461,$ (d) $k=0.4701.$}
\end{figure}

Our results are expected to be consistent with Rabi theory when RWA works. In the terms used here it corresponds to the approximate resonance condition $\nu=\omega_0$ ($\Delta=1$). For $N=1$ under this relation the expression for the $S_3(\tau)$ can be simplified
\[S_3(\tau)=\sqrt{1-k^2\textmd{sn}^4\tau}\cos\left[k\tau+\frac{\imath}{2}\ln\left(
\frac{\textmd{dn}\tau-\imath k\textmd{sn}\tau\textmd{cn}\tau}
{\textmd{dn}\tau+\imath k\textmd{sn}\tau\textmd{cn}\tau}\right)\right]\]
In the corresponding Rabi field ${\bf B}_R$
dynamics is given by $S_3(\tau)=\cos k\tau,$ as $a=k.$ Therefore, simultaneous satisfaction of $\textmd{sn}(\tau_{0},k)=0$ and $\cos(k \tau_{0})=1$ ($\tau_{0}=2mK(k),~m \in \mathbb{Z}$ and $\tau_{0}=\pi(2n+1)/k,~n \in \mathbb{Z},$ respectively) gives values of $k$ corresponding to Rabi resonance.
The ratio $m/(2n+1)$ have to be an integer, we set $n=0$ for brevity. Using the known relation between the incomplete elliptic integral of the first kind $K(k)$ and the hypergeometric function \cite{WW}, the resonance condition for $N=1$ takes the form $k~_{2}F_1(1/2,1/2;1;k^2)=1/m.$ It can be generalized for the arbitrary $N,$ meaning that
\emph{Rabi resonance} occurs under \begin{equation}\label{genrescond}
k~_{2}F_1(1/2,1/2;1;k^2)=\frac{1}{mN}.\end{equation}

The solutions presented here under resonance conditions $\nu=\omega_0$ and (\ref{genrescond}) are consistent with the results of the numerical integration of the problem with linearly polarized harmonic wave of the same amplitude and frequency. As well, they resemble dynamics predicted under the RWA (Fig.~\ref{pict2}).
If one is interested to study dynamics of the two-level system induced by the linearly polarized harmonic field in the extended range of amplitudes of time-dependent field, it is necessary to calculate solutions with higher $N$ due to the proposed scheme. As the amplitude is $Nk,$  the increase of $N$ also allows to decrease $k$ making the approximation better.

If under a two-level system the spin 1/2 system is regarded, the results obtained here can be easily generalized to the spin $j$ system (equivalently, to $2 j+1$ non-interacting spins 1/2). The amplitudes $C_\pm(t)$ with respect to the initial conditions (\ref{incond})
form the matrix of evolution $\hat{u}(t)\in \textmd{SU}(2),$ with the elements $\alpha=C_+,~\beta=-C_-^*.$ With known $\hat{u}(t)$ one can get the solution for the spin $j$ system due to the well-known procedure (see for detail \cite{Bezvershenko}).

\section{Conclusions}
In this work we have considered the problem of two-level system dynamics induced by the external time-dependent field, expressed in terms of the Jacobi elliptic function (cnoid). We presented the scheme for constructing the exact solutions of the Schr\"odinger equation with such a field and used it to calculate explicit expressions for the amplitudes. At different parameters the chosen field corresponds to three physical situations: a nonlinearly modulated Rabi (rotating) field, with $k=1$ it is the N-soliton pulse and with $\omega=0$ and small enough $k$ it resembles linearly polarized harmonic wave. For all these field configurations we discussed the resonance conditions. In particular, in the case of linearly polarized field the resonance condition in the analytic form reproduce at small field amplitude the Bloch-Siegert shift. The explicit solutions for $N=1,~2$ with $\omega=0$ are shown to coincide at small field amplitudes and near resonance with predictions of the rotating wave approximation and to be in good agreement with numerical calculations in the wide range of field parameters beyond it.

\section*{Acknowledgements}

We are grateful to V.Z. Enol'skii, D.V. Leikin and E.D. Belokolos for engaged discussions.
This work  is supported partially by the International
Charitable Fund for the Renaissance of Kyiv-Mohyla Academy.



\begin{thebibliography}{99}
\bibitem{RemarkEberly} Such a field configuration with $N=1$ was proposed in \cite{Eberly} for the optical Bloch vector dynamics. But partial solution found there doesn't satisfy proper initial condition for the quantum system.
\end{thebibliography}
\end{document}